\documentclass[aps, 11pt, preprint, superscriptaddress]{revtex4}

\usepackage{CJK}
\usepackage[usenames]{xcolor}
\usepackage{graphicx}
\usepackage{amsmath}
\usepackage{amsfonts}
\usepackage{amssymb}
\usepackage{mathrsfs}
\usepackage{bm}
\usepackage{verbatim}
\usepackage{cancel}
\usepackage{multirow}
\usepackage{commath}
\usepackage{slashed}

\newcommand{\beq}{\begin{equation}}
\newcommand{\eeq}{\end{equation}}
\newcommand{\bea}{\begin{eqnarray}}
\newcommand{\eea}{\end{eqnarray}}

\graphicspath{{../rev/}}

\begin{document}

\title{Triangle diagram in heavy-baryon chiral perturbation theory}

\author{Songlin Lyu}

\author{Bingwei Long}
\email{bingwei@scu.edu.cn}
\affiliation{College of Physics, Sichuan University, 29 Wang-Jiang Road, Chengdu, Sichuan 610064, China}

\preprint{CTP-SCU/2016006}
\preprint{INT-PUB-16-015}

\date{\today}

\begin{abstract}
The pion-baryon triangle diagram is inspected for a special kinematic region where the squared momentum transfer $t$ is close to $4m_\pi^2$: $|t - 4m_\pi^2| \lesssim m_\pi^4/m_N^2$. Instead of arguing on the ground of anomalous threshold, we investigate possible impacts on power counting. The pion can have very small energies, as opposed to $\sim m_\pi$ in the physical region, which allows all three propagators to be extremely near their mass shell and contributes significantly to the loop integral. We find that in this narrow kinematic window the static-limit approximation for the baryon propagator is invalid and the resummation of the kinetic energy is necessary. Interestingly, in contrast to low-energy two-baryon processes, this resummation of baryon recoils does not lead to overall enhancement of power counting of the diagram.
\end{abstract}

\maketitle

\section{Introduction\label{sec_intro}}

\begin{figure}
    \centering
    \includegraphics[scale=0.8]{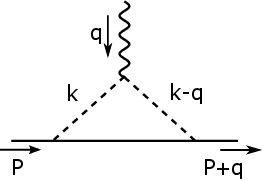}
    \caption{The triangle diagram analyzed in the present paper. The solid (dashed) line represents the baryon (pion). The wavy line represents possible probes allowed by symmetries.}
    \label{fig_trangle}
\end{figure}

As far as one-baryon processes are concerned, baryons have always been approximated as static objects at leading order in heavy-baryon chiral perturbation theory (HBChPT), and recoil corrections (or kinetic energies) are treated as subleading perturbations~\cite{Jenkins_1990jv}:
\begin{equation}
\frac{1}{p_0 -\frac{\vec{p}\,^2}{2m_N} + \frac{\vec{p}\,^4}{8m_N^3} + \cdots} = \frac{1}{p_0} + \frac{\vec{p}\,^2}{2 m_N p_0} + \cdots \, ,
\end{equation}
where $p_\mu = (p_0+m_N, \vec{p}\,)$ is the four-momentum flowing through the baryon propagator and $m_N$ is the baryon mass. This point of view, however, is challenged by the phenomenological successes of covariant approaches towards ChPT, in which recoil corrections of the baryon are in effect resummed~\cite{Becher_1999he, Fuchs_2003qc, Pascalutsa_2002pi, Lensky_2014dda, Alarcon_2011zs, Chen_2012nx, Geng_2008mf, MartinCamalich_2010fp, Ren_2012aj, Geng_2013xn}. The first theoretical rationale for covariant treatment of the baryon comes from Ref.~\cite{Becher_1999he} (also touched upon in Refs.~\cite{Gasser_1987rb, Bernard_1996cc}), in which the pion-baryon triangle diagram was examined in depth. Figure~\ref{fig_trangle} shows the diagram under consideration, where $p$ is the incoming 4-momentum of the baryon and $q$ is the momentum transfer. With different vertexes, this diagram contributes to various processes. In particular, when both baryonic external lines are on-shell the diagram contributes to most baryon form factors, and the loop integral is a function of $t \equiv q^2 = q_0^2 - \vec{q}\,^2$.

If the baryon propagator is fully Lorentz covariant, the triangle diagram can be shown to have a branch-point singularity in the second Riemann sheet of $t$, an example of the so-called anomalous threshold~\cite{Nambu_1957, Landau_1959, Fowler_1960, Eden_Smatrix}:
\begin{equation}
  t = 4m_\pi^2-\frac{m_\pi^4}{m_N^2} \, ,
\end{equation}
where $m_\pi$ is the pion mass. As a direct consequence of this singularity, an anomalously large enhancement around the threshold, was recently used to explain the hadron spectroscopy, especially the resonance-like structures in the invariant mass spectra~\cite{Liu:2015taa,Bayar:2016ftu,Wu:2011yx}. However, applications to these hadronic phenomena are not the focus of the present paper; we are instead interested in power counting for the particular system of pion and baryon where three-momenta are comparable to $m_\pi$ but much smaller than the baryon mass.

Reference~\cite{Becher_1999he} argued that the static-limit approximation is not aware of this second-sheet singularity, so a covariant treatment is necessary. However, from a somewhat puristic point of view towards effective field theory (EFT), symmetries and degrees of freedom are the only constraints one starts with. Whenever resummation of a certain class of terms is deemed necessary, it must come from power counting. Along this line of thinking, if the anomalous threshold in the triangle diagram does turn out to account for important physics, one would like such an analytic structure to emerge from power counting, not the other way around. In doing so, one may hope that the insight gained from power counting could be applied to other processes.

Before more detailed explanation of our findings in next sections, we would like to draw a rough outline about the physics so that it would not get lost in the rather mathematical presentation.
Our focus is a small kinematic window around the two-pion branch point $t = 4 m_\pi^2$.
Integrating out the zeroth component of the pion momentum $k$, as defined in Fig.~\ref{fig_trangle}, we can analytically continue the integral into such an unphysical region by deforming the hypercontour of $\vec{k}$ onto the complex domain where $\vec{k}^2$ takes negative values. In turn, the energy of the baryon propagator, $p_0 + (\vec{k}^2 + m^2_\pi)^{\frac{1}{2}}$, can take values much smaller than $m_\pi$ because of the cancellation between $\vec{k}^2$ and $m_\pi^2$. This immediately jeopardizes the static limit usually adopted as first approximation in HBChPT, because $(\vec{k}^2 + m^2_\pi)^{\frac{1}{2}}$ can now be suppressed as much as the baryon recoil $\vec{k}^2/2m_N$.

\begin{figure}
    \centering
    \includegraphics[scale=0.7]{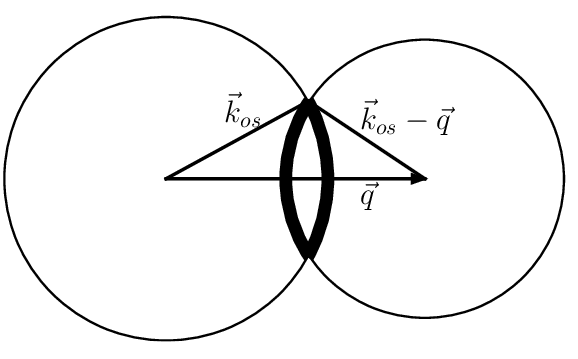}
    \caption{``Dual'' diagram of three-momenta in which vectors may take complex values. Represented by thick arcs are the integration regions where cancellation between $\vec{k}^2$ and $m_\pi^2$ gives $(\vec{k}^2 + m_\pi^2 )^\frac{1}{2} \sim m_\pi^2/m_N$.
    Here $\vec{k}$ is understood as complex vector and $\vec{k}_{os}$ is defined in Eq.~\eqref{eq_pionshell}.}
    \label{fig_sphere}
\end{figure}

Moreover, the pertinent integration domain of $\vec{k}$ can be illustrated in Fig.~\ref{fig_sphere}, where the said cancellation happens to $(\vec{k}^2 + m^2_\pi)^{\frac{1}{2}}$: the domes defined by two intersected spheres. Although the volume is tiny, all three propagators in the loop are extremely close to their mass shell. The end result is that we need to resum the recoil term of the baryon propagator in a small domain centered around $t=4m_\pi^2$. This will be explained in detail in Sec.~\ref{sec_recoil}.

The near-two-pion-cut kinematics poses an interesting comparison with two nucleons interacting at low momenta, $Q \ll m_N$, where $Q$ refers generically to the size of external momenta. There, the kinetic energy $\vec{k}^2/2m_N$ is also included in the nucleon propagator at leading order (LO), as opposed to the static-limit approximation~\cite{Weinberg_1991um}. The stark contrast is that the two-nucleon diagrams are enhanced by $\mathcal{O}(m_N/m_\pi)$, whereas the triangle diagram is not enhanced despite the resummation of the kinetic energy, as will be discussed in Sec.~\ref{sec_recoil}.

The manuscript is structured as follows. We first give a qualitative analysis and point out the configurations where the recoil term needs to be resummed. Numerical results are shown in Sec.~\ref{sec_numerics} to substantiate the argumentation, followed by a summary and conclusion in Sec.~\ref{sec_discussion}.

\section{Kinematics around two-pion cut \label{sec_recoil}}

We choose the rest frame of the incoming baryon, so that its four-momentum is $(m_N, \vec{0}\,)$ where $m_N$ is the baryon mass. The outgoing baryon has four-momentum $(m_N+q_0, \vec{q}\,)$, and the baryon lines being on-shell imposes the following constraint on $q_0$ and $\vec{q}\,^2$ as functions of $t = q_0^2 - \vec{q}\,^2$:
\begin{equation}
    q_0(t) = -\frac{t}{2m_N}\, , \qquad \vec{q}\,^2(t) = t\left(-1 + \frac{t}{4m_N^2}\right) \, .
    \label{eqn_q0qvec}
\end{equation}
In the physical region, $q_0$ and $\vec{q}\,^2$ are both positive real numbers; hence, $-4 m_N^2 < t < 0 $. ChPT is expected to work in regions in the complex $t$-plane where $|\vec{q}\,|$ is sufficiently small. In addition, the vertexes are not crucial for the purpose of analyzing power counting.


The loop integral being investigated is given by
\begin{equation}
  \gamma(t, m_\pi^2) \equiv i \int \frac{\mathrm{d}^4 k}{(2\pi)^4} \frac{1}{k_0 - \frac{\vec{k}^2}{2m_N} + \cdots + i \epsilon} \frac{1}{k^2 - m_\pi^2 + i\epsilon} \frac{1}{(k - q)^2 - m_\pi^2 + i\epsilon} \, , \label{eqn_gamma}
\end{equation}
where $q$ and $k$ label momenta as depicted in Fig.~\ref{fig_trangle}. The baryon propagator is not fully covariant: it propagates only forward with time, which is the common premise in HBChPT. But only for the time being, we have retained the kinematic corrections in $\vec{k}^2/m_N^2$, delaying expansion until the power counting becomes clear. Note that $\gamma(t)$ defined by the above integral differs from $\gamma(t)$ defined by Eq.(4) in Ref.~\cite{Becher_1999he}, where the baryon propagator is Lorentz covariant: $(k^2 - m_N^2)^{-1}$.

It is often more reliable to power count with only three-momenta remaining. Performing the $k_0$ integral gives two residues of two pions poles
\begin{equation}
\begin{split}
  \gamma_1 &= - \int \frac{\mathrm{d}^3 k}{(2\pi)^3} \frac{1}{\sqrt{\vec{k}^2 + m_\pi^2} + \frac{\vec{k}^2}{2m_N} + \cdots} \\
  & \quad \quad \times \frac{1}{2
  \sqrt{\vec{k}^2 + m_\pi^2}}\, \frac{1}{\big(q_0 + \sqrt{\vec{k}^2 + m_\pi^2} \, \big)^2 - (\vec{k} - \vec{q} \,)^2 - m_\pi^2}\, , \label{eqn_gamma1}
\end{split}
\end{equation}
and
\begin{equation}
\begin{split}
  \gamma_2 &= \int \frac{\mathrm{d}^3 k}{(2\pi)^3} \frac{1}{q_0 - \sqrt{(\vec{k} - \vec{q} \,)^2 + m_\pi^2} - \frac{\vec{k}^2}{2m_N^2} + \cdots} \\
  & \quad \quad \times \frac{1}{2 \sqrt{(\vec{k} - \vec{q} \,)^2 + m_\pi^2}}\, \frac{1}{\big(q_0 - \sqrt{(\vec{k} - \vec{q} \,)^2 + m_\pi^2} \, \big)^2 - \vec{k}^2 - m_\pi^2}\, . \label{eqn_gamma2}
\end{split}
\end{equation}
In doing so we have made one of the pions always be on-mass-shell.

When power counting loop diagrams \`a la Weinberg~\cite{Weinberg_1978kz}, one tries to capture long-range physics, by inspecting contributions from loop momenta that push at least a subset of the propagators close to their mass shell. The off-shellness of those propagators must be comparable in size to external kinematic variables like $\vec{q}\,^2$ and/or $m_\pi^2$. It is assumed throughout the paper $\vec{q}\,^2 = \mathcal{O}(m_\pi^2)$.

We focus on $\gamma_1$ first. When both pion propagators are on-mass-shell, the loop momentum $\vec{k}_{os}$ satisfies the following equation:
\begin{equation}
  \vec{q}\,^2 - 2\vec{k}_{os} \cdot \vec{q}  = 2q_0\sqrt{\vec{k}_{os}^2 + m_\pi^2} + q_0^2 \, , \label{eq_pionshell}
\end{equation}
which also sets the pole position of the second term in the second line of Eq.~\eqref{eqn_gamma1}. Using $q_0 \sim \vec{q}\,^2/2m_N$, one finds that $\vec{k}_{os}$ is roughly $\vec{q}/2$:
\begin{equation}
  \vec{k}_{os} = \frac{\vec{q}}{2} + \mathcal{O}\left(\frac{\vec{q}\,^2}{m_N}\right) \, . \label{kos_qvec}
\end{equation}
In the physical region where $t < 0$ and $\vec{q}$ is real, the hypercontour of $\vec{k}$ stays largely in the real domain except for an infinitesimal detour to circumvent $\vec{k}_{os}$. This is important because there will \emph{not} be cancellation between $\vec{k}^2$ and $m_\pi^2$ in the propagators: $\vec{k}^2 + m_\pi^2 \sim m_\pi^2$. It follows immediately that the denominator of the baryon propagator,
\begin{equation}
  (\vec{k}^2 + m_\pi^2)^\frac{1}{2} + \frac{\vec{k}^2}{2m_N} + \cdots \, ,
\end{equation}
is dominated by the pion energy $(\vec{k}^2 + m_\pi^2)^\frac{1}{2} \sim m_\pi$, whereas the recoil correction $\vec{k}^2/2m_N \sim \xi m_\pi$ is subleading, where $\xi \equiv m_\pi/m_N$. Therefore, the static limit is justified and recoil corrections are all dropped in first approximation. With the baryon propagator and the two pion propagators being counted respectively as $m_\pi^{-1}$, $m_\pi^{-1}$, and $m_\pi^{-2}$, one concludes that $\gamma_1$ is of order $m_\pi^{-1}$. $\gamma_2$ in Eq.~\eqref{eqn_gamma2} is similarly counted, so we arrive at the standard counting that the loop integral $\gamma(t, m_\pi^2)$ is of order $m_\pi^{-1}$ in the physical region where $\vec{q}\,^2 > 0 $ and $\vec{q}\,^2 = \mathcal{O}(m_\pi^2)$.

While the above argument covers the generic case of $\vec{q}\,^2 \sim m_\pi^2$, it, however, needs modification for peculiar cases defined by special values of $\vec{q}\,^2$. Pertinent to our concern in the present paper is a very small region around the two-pion cut:
\begin{equation}
  t = 4m_\pi^2  \quad \text{and} \quad \vec{q}\,^2 = -4m_\pi^2 + \mathcal{O}(\xi^2 m_\pi^2) \, , \label{qvec_reg}
\end{equation}
where both pions are near their mass shell simultaneously.

\begin{figure}
    \centering
    \includegraphics[scale=0.6]{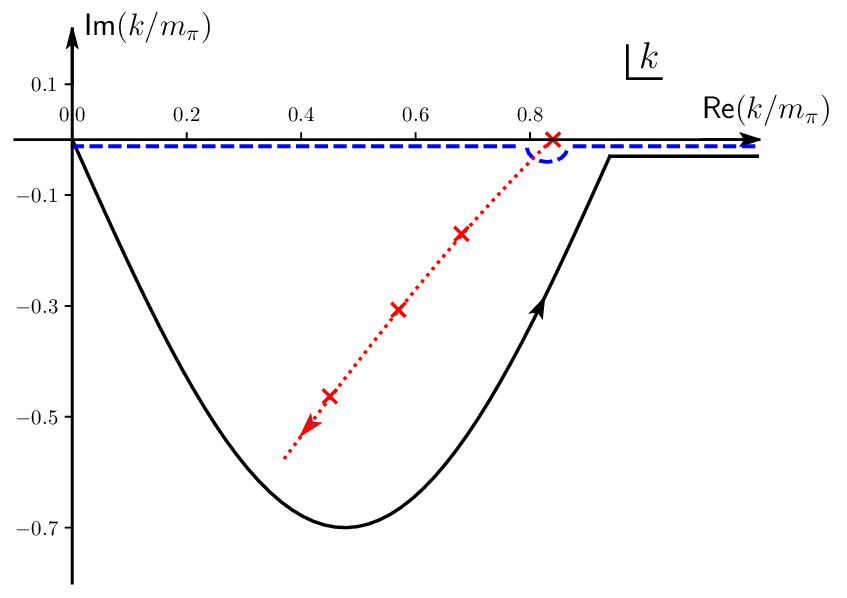}
    \caption{The $k$ contours and poles corresponding to $\vec{k}_{os}$. The dotted line with crossed markers represents movement of the pole and the solid (dashed) line illustrates the deformed (undeformed) contour. See the text for detailed explanation.}
    \label{fig_pole_kcos}
\end{figure}

This unphysical region can be accessed through contour deformation of the momentum integrations.
We illustrate the deformed contour with the integral of $\gamma_1$. Integrating out one of the angles in $\vec{k}$ space, one is left with integration over
\begin{equation*}
  k \equiv |\vec{k}\,| \quad \text{and} \quad \cos\theta \equiv \frac{\vec{k} \cdot \vec{q}}{\sqrt{\vec{k}^2\,\vec{q}\,^2}} \, .
\end{equation*}
By choice, we always integrate along the real axis of $\cos \theta$ from $-1$ to $1$. For fixed $\cos \theta$, the undeformed contour in the $k$ plane goes from the origin to $+\infty$, as represented by the dashed line in Fig.~\ref{fig_pole_kcos}. Again, with fixed $\cos \theta$, $\vec{k}_{os}$ is represented by the pole of the integrand along the positive real axis, and it intends to cross the contour when $\vec{q}\,^2$ obtains an imaginary part. To define a continuous integral, we can deform the $k$ contour to avoid the crossing. Doing so makes $k$ acquire complex values. Figure~\ref{fig_pole_kcos} shows the movement of the pole and the according deformation of the contour, when $\vec{q}\,^2/m_\pi^2$ varies from $4.41$ to $-1.26 - 2.03i$ and $\cos\theta$ is fixed at $0.98$, .

For most of $\vec{k}$ space, one can still use the static-limit approximation for the baryon propagator. But there is a small integration domain in which the recoil term needs to be resummed. For the integral of $\gamma_1$ [Eq.~\eqref{eqn_gamma1}], consider the following region of $\vec{k}$ space around $\vec{k}_{os}$:
\begin{equation}
  \vec{k}\,^2 = -m_\pi^2 + \mathcal{O}(\xi^2 m_\pi^2) \, ,
\end{equation}
such that the cancellation between $\vec{k}^2$ and $m_\pi^2$ gives
\begin{equation}
  \sqrt{\vec{k}^2 + m_\pi^2} = \mathcal{O}(\xi m_\pi) \quad \mathrm{and} \quad 2\vec{k}\cdot\vec{q} - \vec{q}\,^2 =\mathcal{O}(\xi^2 m_\pi^2) \, , \label{gamma1_cond}
\end{equation}
where we have used $\vec{q}\,^2\simeq-4m_\pi^2$. The above constraints define a region that looks like a dome centered at the origin with radius $i m_\pi$, illustrated in Fig.~\ref{fig_sphere} as the thick arc on the right.
For the points on the dome, we can write
\begin{equation*}
  \vec{k} = \left(i m_\pi +\Delta k\right) \hat{\vec{k}} \, ,
\end{equation*}
where $\hat{\vec{k}}$ is the unit vector along $\vec{k}$. Therefore, it follows from the the first constraint on the dome in Eq.~\eqref{gamma1_cond} that
\begin{equation}
	\sqrt{2i m_\pi \Delta k} = \mathcal{O}(\xi m_\pi) \,,
\end{equation}
i.e., the thickness of the spherical shell is $\Delta k = \mathcal{O}(\xi^2 m_\pi)$. From the second constraint in Eq.~\eqref{gamma1_cond}, we can obtain that $\vec{k}$ must be almost collinear with $\vec{q}$ to ensure
\begin{equation}
	2\vec{k}\cdot\vec{q} \simeq -4m_\pi^2 \left(1-\frac{\theta^2}{2} + \mathcal{O}(\theta^4)\right) = -4m_\pi^2 + \mathcal{O}(\xi^2m_\pi^2) \,,
\end{equation}
and $\theta \sim \mathcal{O}(\xi)$ is the deviation angle between $\vec{k}$ and $\vec{q}$, which is the polar angle of the dome. Therefore, the dome subtends an angle of order $2\theta \sim \mathcal{O}(\xi)$. The solid angle of this dome is $\Omega \sim \mathcal{O}(\pi\xi^2)$, so the integration volume is counted as follows:
\begin{equation}
    \int \mathrm{d}^3 k \; = \; \int \mathrm{d} k\, \mathrm{d}\Omega\, |\vec{k}^2|  \; \sim \; m_\pi^2 \xi^2 m_\pi \xi^2 \; \sim \; \xi^4 m_\pi^3 \,.
\end{equation}

On the other hand, all of the three propagators are significantly enhanced because they are extremely close to their mass shell and they combine to contribute significantly despite the smallness of the integration volume.

The recoil term of the baryon propagator is as important as the pion energy,
\begin{equation}
  \frac{\vec{k}^2}{2m_N} \sim \sqrt{\vec{k}^2+m_\pi^2} = \mathcal{O}(\xi m_\pi) \, ;
\end{equation}
therefore, it can no longer be considered subleading and the static limit is consequently invalidated. It follows that the baryon propagator in Eq.~\eqref{eqn_gamma1} is enhanced as
\begin{equation}
  \frac{1}{\sqrt{\vec{k}^2 + m_\pi^2} + \frac{\vec{k}^2}{2m_N}}\sim \frac{1}{\xi m_\pi} \, ,
\end{equation}
where we have dropped $\vec{k}\,^4/m_N^3$ and higher-order terms. Using Eq.~\eqref{gamma1_cond} and $q_0\sim \xi m_\pi$, we find that the pion propagators in Eq.~\eqref{eqn_gamma1} are enhanced too:
\begin{align}
    \frac{1}{\sqrt{\vec{k}^2 + m_\pi^2}}  \, & \, \sim \xi^{-1} m_\pi^{-1} \,, \label{scale_gamma1pion1} \\
    \frac{1}{q_0^2 + 2q_0\sqrt{\vec{k}^2 + m_\pi^2} - \vec{q}\,^2 + 2\vec{k}\cdot\vec{q}} \, \, & \, \sim \xi^{-2} m_\pi^{-2} \, . \label{scale_gamma1pion2}
\end{align}
The end result is that in this particular integration domain $\gamma_1$ in Eq.~\eqref{eqn_gamma1} scales as
\begin{equation}
  \gamma_1 \sim \xi^4 m_\pi^3 \, \frac{1}{\xi m_\pi} \, \frac{1}{\xi m_\pi} \, \frac{1}{(\xi m_\pi)^2} \sim \frac{1}{m_\pi} \, .
\end{equation}
The key here is that the enhancement of the propagators makes up for the smallness of the integration measure, to the extent that they combine to contribute comparably as the integration outside the dome, both of order $\sim m_\pi^{-1}$. So we arrived at the conclusion that $\gamma_1 \sim m_\pi^{-1}$ in this unphysical kinematic region.

For $\gamma_2$ defined in Eq.~\eqref{eqn_gamma2}, similar enhancement of propagators happens, but in different domain of $\vec{k}$, illustrated as the thick arc on the left in Fig.~\ref{fig_sphere}. It is then straightforward to obtain that $\gamma_2 \sim m_\pi^{-1}$. Finally, we conclude that the loop integral $\gamma(t, m_\pi^2)$ is of $\mathcal{O}(m_\pi^{-1})$.

To summarize, we find that when $t$ is within a small window around $4m_\pi^2$, $|t - 4m_\pi^2| \lesssim \xi^2 m_\pi^2$, the static limit is no longer a valid approximation: the recoil term $-\vec{k}^2/2m_N$ must be retained in Eq.~\eqref{eqn_gamma}. However , it must be stressed that the resummation of $\vec{k}^2/2m_N$ does not change the overall counting of the integral: $\gamma \sim m_\pi^{-1}$ as it is counted when $t$ is outside the said window.

\section{Numerics\label{sec_numerics}}

We evaluate numerically the integral \eqref{eqn_gamma} with only the first recoil term kept in the baryon propagator
\begin{equation}
    \gamma^{(0)}(t, m_\pi^2) = i \int \frac{\mathrm{d}^4 k}{(2\pi)^4} \frac{1}{k_0 - \frac{\vec{k}^2}{2m_N} + i \epsilon} \frac{1}{k^2 - m_\pi^2 + i\epsilon} \frac{1}{(k - q)^2 - m_\pi^2 + i\epsilon} \,.
\end{equation}
A Feynman parameter is first used to combine the pion propagators and then the following identity is used to incorporate the baryon propagator:
\begin{equation}
  \frac{1}{ab} = 2 \int_0^\infty \mathrm{d} \lambda \frac{1}{\left(a + 2\lambda b \right)^2} \, .
\end{equation}
We then proceed by integrating out $k$ after completing the square in $k$, eventually arriving at
\begin{equation}
\begin{split}
    \gamma^{(0)}(t, m_\pi^2) &= \frac{1}{8\pi^2} \int_0^1 \mathrm{d}x \int_0^{\infty} \frac{\mathrm{d}\lambda}{\sqrt{1+\frac{\lambda}{m_N}}} \\
    & \qquad \times \frac{1}{x^2\left(q_0^2 - \frac{\vec{q}\,^2}{1 + \lambda/m_N}\right) - x\left(q_0^2 - \vec{q}\,^2 + 2\lambda q_0\right) + m_\pi^2 + \lambda^2} \, . \label{eqn_doubleintegral}
\end{split}
\end{equation}

We wish to compare the above integral with its relativistic version:
\begin{equation}
  \gamma_{rel}(t, m_\pi^2) = i \int \frac{\mathrm{d}^4 k}{(2\pi)^4} \frac{2m_N}{k^2 - m_N^2 + i \epsilon} \frac{1}{k^2 - m_\pi^2 + i\epsilon} \frac{1}{(k - q)^2 - m_\pi^2 + i\epsilon} \, , \label{eqn_relgamma}
\end{equation}
which has a well-known branch point at $t = 4m_\pi^2$. The comparison is facilitated if the branch point of $\gamma^{(0)}(t, m_\pi^2)$ coincides that of $\gamma_{rel}(t, m_\pi^2)$. This can be achieved by retaining $q_0(t)$ and $\vec{q}\,^2(t)$ as defined in Eq.~\eqref{eqn_q0qvec}. If we had chosen to conform to non-relativistic kinematics for the outgoing baryon, $q_0 = \vec{q}\,^2/2m_N$, $q_0(t)$ and $\vec{q}\,^2(t)$ would have been
\begin{equation}
  q_0(t) = - \frac{t}{m_N\left(1 + \sqrt{1 + \frac{t}{m_N^2}}\right)} \, , \qquad \vec{q}\,^2(t) = - \frac{2t}{\left(1 + \sqrt{1 + \frac{t}{m_N^2}}\right)} \, . \label{eqn_nonrelq0qvec}
\end{equation}
However, the discrepancy between Eqs.~\eqref{eqn_q0qvec} and \eqref{eqn_nonrelq0qvec} is not crucial for our qualitative statement regarding power counting.

To see how the branch point arises at $t = 4m_\pi^2$, we first notice that when $\lambda = 0$ the integrand has two poles in the $x$ plane, and they will pinch at $x = 1/2$ when $t = 4m_\pi^2$. Because the pinching in the $x$ plane occurs at one of the end points of the $\lambda$ integration, the corresponding singularity of the integrand is inevitable no matter how we deform the contours of $\lambda$ and/or $x$. Therefore, $\gamma^{(0)}(t, m_\pi^2)$ has a branch cut starting from $t = 4m_\pi^2$, running toward $+\infty$ along the positive real axis.

\begin{figure}
  \centering
  \includegraphics[scale=0.6]{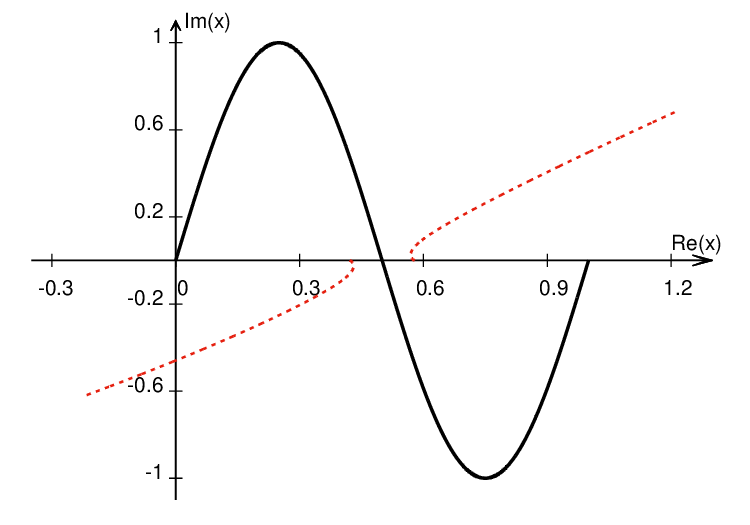}
  \hspace{5mm}
  \includegraphics[scale=0.6]{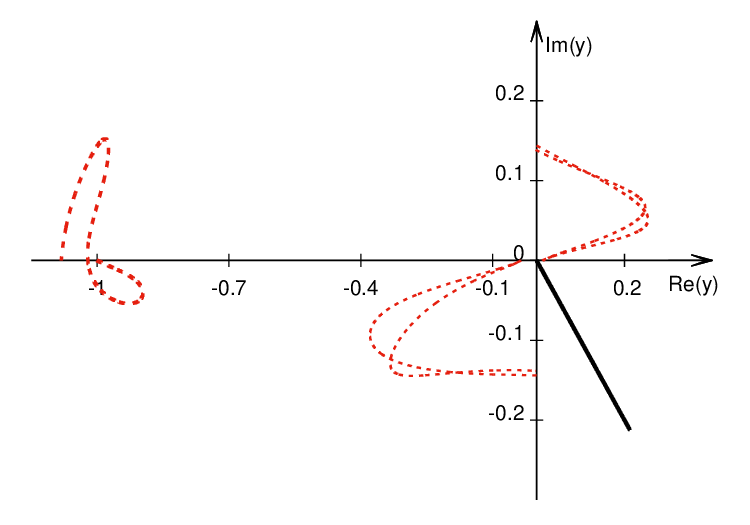}
  \caption{The solid lines illustrate the contours of $x$ and $\lambda$ integrations, where $y \equiv \lambda/m_N$. The dashed lines represent the poles of the integrand in Eq.~\eqref{eqn_doubleintegral} when $t = 4.1 m_\pi^2$ (see the text for more detailed explanation).}
  \label{fig_xypoles}
\end{figure}

For $t > 4m_\pi^2$, the contours in both $x$ and $\lambda$ planes need to be deformed so as not to be crossed by the poles of the integrand. The chosen contours are shown by the solid lines in Fig.~\ref{fig_xypoles}. In particular, the $\lambda$ contour is a straight line from the origin to infinity, $45^\circ$ off the positive real axis. When $\lambda$ takes a value on that contour, the integrand in Eq.~\eqref{eqn_doubleintegral} has corresponding poles in the $x$ plane. The trajectories of these poles as $\lambda$ moving along its contour are represented by the dashed lines in Fig.~\ref{fig_xypoles} (a), with $t = 4.1 m_\pi^2$. A similar story goes to Fig.~\ref{fig_xypoles} (b), in which the $\lambda$ contour is illustrated by the solid line and the poles associated with the $x$ contour are marked out by the dashed lines. On a side note, there are other ways to evaluate the integral. For example, one can first calculate the imaginary part, i.e., the discontinuity along the branch cut, and then the real part by way of dispersion integral. In fact, this is how $\gamma_{rel}$ [Eq.~\eqref{eqn_relgamma}] is evaluated.

\begin{figure}
  \centering
  \includegraphics[scale=0.6]{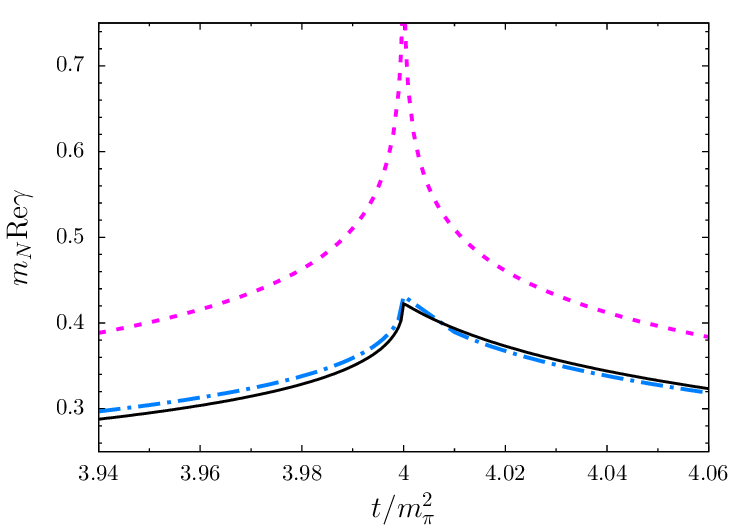}
  \hspace{7mm}
  \includegraphics[scale=0.6]{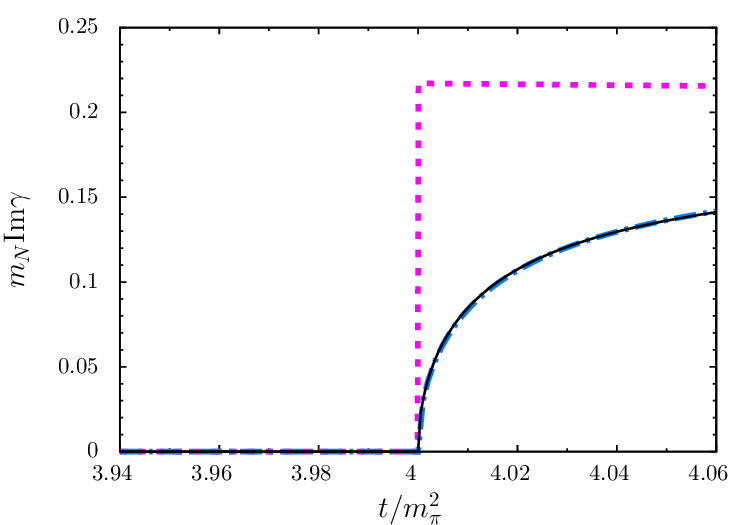}
  \caption{Real and imaginary parts of $\gamma^{(0)}(t, m_\pi^2)$ in neighborhood of $t = 4m_\pi^2$. The dashed, dot-dashed, and solid lines correspond respectively to calculations in the static limit, with the relativistic baryon propagator [Eq.~\eqref{eqn_relgamma}], and with the first recoil term [Eq.~\eqref{eqn_doubleintegral}]. The real part of the static approximation diverges at $t = 4m_\pi^2$.}
  \label{fig_gamma_smallregion}
\end{figure}

The numerical results for the real and imaginary parts of $\gamma^{(0)}(t, m_\pi^2)$ are plotted in Fig.~\ref{fig_gamma_smallregion}, with $m_\pi/m_N = 0.149$--- the ratio between the physical pion and nucleon masses, inside the interested $t$ window
\begin{equation}
  \left|\frac{t}{m_\pi^2} - 4 \right| \lesssim \xi^2  = 0.022 \, .
\end{equation}
The solid lines correspond to $\gamma^{(0)}(t, m_\pi^2)$ [Eq.~\ref{eqn_doubleintegral}]. For comparison, we have also plotted the static-limit approximation (dashed):
\begin{equation}
  \gamma_{static}(t) \; = \;
  \begin{cases}
    \dfrac{1}{16\pi \sqrt{t}}\ln \dfrac{2m_{\pi}+\sqrt{t}}{2m_{\pi}-\sqrt{t}} \,, & 0<t<4m_\pi\,, \\[3ex]
    \dfrac{1}{16\pi \sqrt{t}} \left[ \ln \dfrac{\sqrt{t}+2m_{\pi}}{\sqrt{t}-2m_{\pi}}+i\pi \right] \,, & t>4m_\pi  \,,
  \end{cases}
\end{equation}
which diverges at $t = 4m_\pi^2$, and the relativistic results (dot-dashed) according to Eq.~\eqref{eqn_relgamma}. The difference between $\gamma^{(0)}$ and relativistic representation $\gamma_{rel}$ is very small, which indicates that the higher-order recoil corrections like $\vec{k}^4/8m_N^3$ are indeed higher order.

\begin{figure}
  \centering
  \includegraphics[scale=0.6]{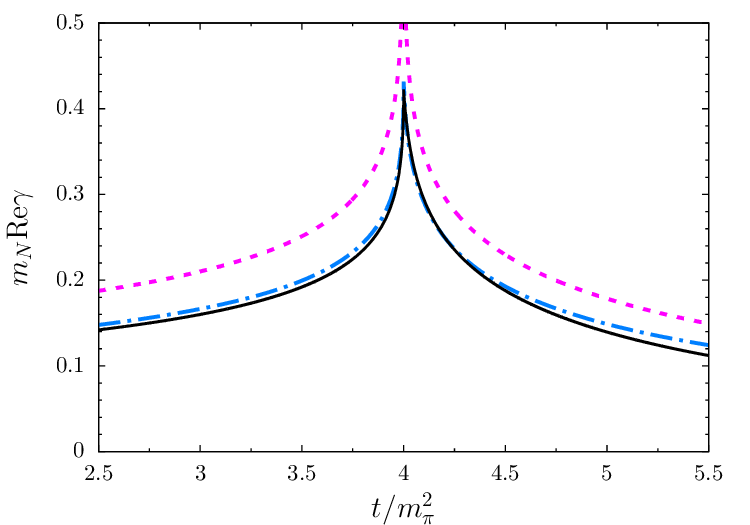}
  \hspace{7mm}
  \includegraphics[scale=0.6]{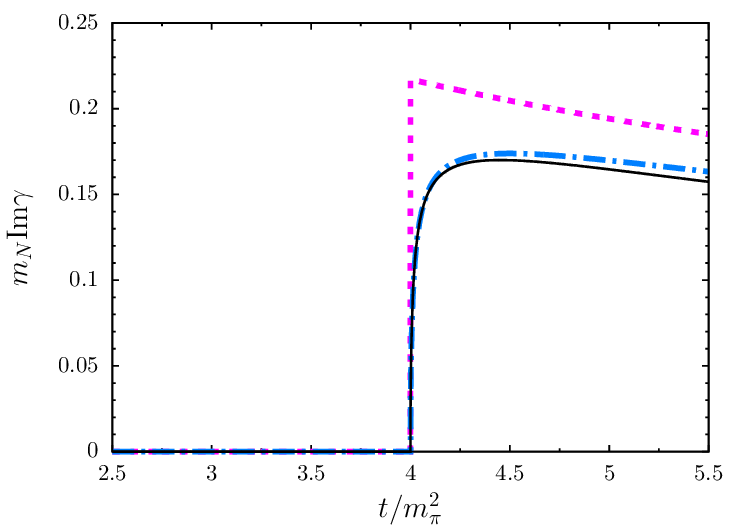}
  \caption{Real and imaginary parts of $\gamma^{(0)}(t, m_\pi^2)$ in a large neighborhood of $t = 4m_\pi^2$.}
  \label{fig_gamma_bigregion}
\end{figure}

Outside this kinematic window, however, the static-limit approximation is still applicable. Shown in Fig.~\ref{fig_gamma_bigregion} is essentially Fig.~\ref{fig_gamma_smallregion} zoomed out, for wider range of t: $2.5 < t/m_\pi^2 < 5.5$. The static limit differs from $\gamma^{(0)}$ by about one third of the latter, presenting acceptable first approximation.

\section{Summary and conclusion\label{sec_discussion}}

We have studied the pion-baryon triangle diagram, which turns out to be yet another example where the recoil term needs to be resummed. More specifically, when the squared momentum transfer $t$ is inside a small window centered on the two-pion cut:
\begin{equation}
  |t - 4m_\pi^2| \lesssim \frac{m_\pi^4}{m_N^2} \, ,
\end{equation}
the first recoil correction $\vec{p}\,^2/2m_N$ must remain in the baryon propagator:
\begin{equation}
  \frac{1}{p_0 - \frac{\vec{p}\,^2}{2m_N} + \frac{\vec{p}\,^4}{8m_N^3} + \cdots} = \frac{1}{p_0 - \frac{\vec{p}\,^2}{2m_N}} - \frac{\vec{p}\,^4/8m_N^3}{\left(p_0 - \frac{\vec{p}\,^2}{2m_N}\right)^2} + \cdots
\end{equation}
However, the integral as a whole is not enhanced, unlike the two-baryon processes at low energies.

In previous studies, the breakdown of the static-limit approximation around the two-pion cut is attributed to its failure to produce the anomalous threshold expected from manifestly covariant triangle diagrams~\cite{Becher_1999he}. In our analysis, we provide more insights, in addition to reproducing the location of the anomalous threshold. The size of the kinematic window in terms of $t$ is estimated, in which the recoil term must be kept. Furthermore, we also show that the contributions of higher-order kinetic-energy terms such as $\vec{k}^4/8m_N^3$ are indeed higher order; therefore, a fully covariant treatment is not crucial, at least as far as power counting is concerned.

\acknowledgments

BwL thanks the nuclear theory group at BeiHang University and the Institute for Nuclear Theory at the University of Washington for hospitality when part of the work was carried out there. This work was supported in part by the National Natural Science Foundation of China (NSFC) under Grant Nos. 11775148 and 11735003, the Fundamental Research Funds for the Central Universities.

\end{document}